\documentclass[numreferences]{kluwer}

\usepackage{amsmath}

\numberwithin{equation}{section}
\numberwithin{equation}{subsection}

\DeclareMathOperator{\Div}{\textbf{Div}}
\DeclareMathOperator{\R}{\textbf{R}}
\DeclareMathOperator{\F}{\textbf{F}}
\DeclareMathOperator{\J}{\textbf{J}}
\DeclareMathOperator{\A}{\textbf{A}}
\DeclareMathOperator{\Pb}{\textbf{P}}

\DeclareMathOperator{\Lb}{\textbf{L}}
\DeclareMathOperator{\g}{\textbf{g}}
\DeclareMathOperator{\Q}{\textbf{Q}}

\newcommand{\delt}{\mbox{\boldmath$\delta$} \hspace{0.1pc}}
\newcommand{\et}{\mbox{\boldmath$\eta$} \hspace{0.1pc}}

\begin{document}

\begin{opening}
\title{Space-time Curvature of Classical Electromagnetism}
\author{R. W. M. \surname{Woodside}\email{Rob.Woodside@UCFV.ca}}
\institute{Physics Dept., University College of the Fraser Valley\\
33844 King Rd., Abbotsford, B.C., Canada, V2S 7M8}

\dedication{Dedicated to the memory of Prof. M. H. L. Pryce
(January 24, 1913--July 24, 2003)}

\runningtitle{Space-time Curvature of Classical Electromagnetism}
\runningauthor{R. W. M. Woodside}

\begin{abstract}

The space-time curvature carried by electromagnetic fields is discovered and 
a new unification
of geometry and electromagnetism is found. Curvature is invariant under 
charge reversal symmetry.
Electromagnetic field equations are examined with De Rham co homology 
theory. Radiative
electromagnetic fields must be exact and co exact to preclude unobserved 
massless topological
charges. Weyl's conformal tensor, here called ``the gravitational field'', 
is decomposed into a
divergence-free non-local piece with support everywhere and a local piece 
with the same support
as the matter. By tuning a local gravitational field to a Maxwell field the 
electromagnetic
field's local gravitational field is discovered. This gravitational field 
carries the
electromagnetic field's polarization or phase information, unlike Maxwell's 
stress-energy tensor.
The unification assumes Einstein's equations and derives Maxwell's equations 
from curvature
assumptions. Gravity forbids magnetic monopoles! This unification is 
stronger than the
Einstein-Maxwell equations alone, as those equations must produce the 
electromagnetic field's
local gravitational field and not just any conformal tensor. Charged black 
holes are examples.
Curvature of radiative null electromagnetic fields is characterized.

\end{abstract}


\begin{ao} \\
R. W. M. Woodside, University College of the Fraser Valley\\
33844 King Rd., Abbotsford, B.C., Canada,
\\V2S 7M8
\end{ao}

\end{opening}

\section{Introduction}

This work discovers the space-time curvature carried by the electromagnetic 
field and provides
a new unification of geometry and classical electromagnetism. The new 
unification contains the
Einstein equations to handle the mechanics and permits the derivation of the 
Maxwell equations
from the full second Bianchi identities. This is a purely classical work and 
quantum
considerations are merely mentioned.

Central to this work are the requirements that the electromagnetic field be 
expressed as a two
form \textbf{F} and fit into general relativity under the demand that the 
total stress-energy
tensor used in the Einstein equations contain the Maxwell stress-energy 
tensor $\mathbf{T_F}$.
In the notation with the conventions of \cite{1} and in S.I. units 
$\mathbf{T_F}$ is
\begin{equation} \label{1.1}
T^{\phantom{F}\alpha}_{F\phantom{\alpha}\beta}=\frac{\varepsilon_{0}}{2}\left(F^{\mu 
\alpha}
F_{\mu \beta}+ *F^{\mu \alpha}*F_{\mu \beta}\right), \end{equation}
where $\varepsilon_{0}=8.85418782\cdot 10^{-12}$ farad/meter is the electric 
vacuum permittivity.

Originally \cite{2} general relativity was conceived as a unification of 
mechanics and geometry
that explained gravitation. It was just a bonus \cite{3} that 
electromagnetism also entered the
unification via equation \eqref{1.1}. If the Maxwell stress-energy tensor 
carried all the
properties of the electromagnetic field, showing electromagnetism to be 
entirely reducible to
mechanics, that would have been the end of the story.

However, the electromagnetic field has
polarization or phase information that is not contained in the Maxwell 
stress-energy tensor
\cite{4}. Since Weyl's conformal tensor, the totally traceless piece of 
Riemann curvature, is
supposed to contain the phase or polarization information carried by 
gravitational radiation,
one should expect it to do the same for electromagnetic radiation.

This is born out by the
discovery of a piece of the Weyl conformal tensor that depends explicitly on 
the
electromagnetic field and contains this polarization or phase information. 
It is denoted by
$\mathbf{C_F}$, called ``the local gravitational field of the 
electromagnetic field'', and
given by:
\begin{equation} \begin{split} \label{1.2}
C^{\phantom{F} \alpha \beta}_{F \phantom{\beta} \gamma \delta}= 
8\pi\frac{G\varepsilon_{0}}{c^4}\{
\frac{3}{2}\left(F^{\alpha \beta}F_{\gamma \delta}-*F^{\alpha 
\beta}*F_{\gamma \delta}\right) +
\\ -\frac{1}{4}\delta^{\alpha \beta}_{\gamma \delta}F^{\mu \nu}F_{\mu 
\nu}+\frac{1}{4}
\eta^{\alpha \beta}_{\phantom{\alpha \beta} \gamma \delta}*F^{\mu \nu}F_{\mu 
\nu}\},
\end{split} \end{equation}
where $G=6.6726\cdot10^{-11}$ Newton-meter$^{2}$/kilogram$^{2}$ is Newton's 
gravitational
constant, $c=2.99792458\cdot10^8$ meter/second is the speed of light, 
$\delt$ is a
fully antisymmetric tensor, and $\et$ is the permutation tensor.
The traces in the expression
\eqref{1.2} for $\mathbf{C_F}$ are the Lorentz invariants of the 
electromagnetic field
\begin{subequations} \label{1.3}
\begin{equation} \label{1.3a}
F^{\mu \nu}F_{\mu \nu}=-2(E^2-c^2B^2)  \end{equation}
and
\begin{equation} \label{1.3b}
*F^{\mu \nu}F_{\mu \nu}=4c\Vec{E} \cdot \Vec{B}, \end{equation}  
\end{subequations}
where $\Vec{E}$ is the electric field strength in Volt/meter and $\Vec{B}$ 
is the magnetic field
strength in Tesla.

The remaining piece of curvature $\mathbf{M_F}$ carried by the 
electromagnetic field is
determined by the Einstein equations and as suggested by equation 
\eqref{1.1} is
\begin{equation} \label{1.4}
M^{\phantom{F} \alpha 
\beta}_{F\phantom{\beta}\gamma\delta}=8\pi\frac{G\varepsilon_{0}}{c^4}\frac{1}{2}
\left(F^{\alpha \beta}F_{\gamma \delta}+ *F^{\alpha \beta} * F_{\gamma 
\delta}\right).
\end{equation}
The space-time curvature $\mathbf{R_F}$ carried by the electromagnetic field 
is then
\begin{equation} \label{1.5}
\mathbf{R_F}=\mathbf{M_F}+\mathbf{C_F}. \end{equation}

The second section of this paper examins various ``electromagnetic'' field 
equations from a
topological
viewpoint.  In addition to the Maxwell equations, this section considers 
other
``electromagnetic'' field equations
consistent with duality rotations \cite{1,5}, classical mechanics, and null 
fields \cite{6}.
It also gives an attempt to derive the Maxwell equations from the Maxwell 
stress-energy
tensor and
Lorentz force density in general relativity, without the benefit of the new 
unification.
This old derivation requires a trivial space-time topology to make all 
closed forms exact and
recover the global vector potential in Maxwell's equations.  The non-trivial 
topology of the
black hole solutions
vitiates this derivation.  However, the new unification precisely selects 
the exact Maxwell
fields. This global vector potential in the Maxwell equations precludes the 
unobserved magnetic
monopoles \cite{5,7,8}.  Thus a major conclusion of this work is that the 
unification of
electromagnetism and geometry also forbids magnetic monopoles.

The second section also finds that ``electromagnetic'' null fields allow the 
possibility
of massless topological magnetic or electric charges that have never been 
observed.
The new unification removes the magnetic charges, but the removal of 
topological massless
electric charges requires the introduction of a second vector potential 
making {\bf F} both exact
and co exact for radiative fields. The Maxwell equations will admit this 
second global vector
potential, but as they stand they also permit massless electric charges.

In the third section on space-time curvature the conformal tensor is 
identified as ``the
gravitational field''. It is invariantly split into a divergence free piece 
called ``the non
local gravitational field'' with support everywhere; and a piece, usually 
with non zero
divergence, called ``the local gravitational field''. Most importantly, the 
local gravitational
field has the same support as the matter. Also the Riemann curvature is 
decomposed by trace and
split with an old technique \cite{9} that is further developed here. This 
greatly simplifies
calculations. It also permits the duality rotation of curvature, which shows 
that curvature is
invariant under charge reversal symmetry. Any quantity explicitly dependant 
on the
electromagnetic field that remains invariant under charge reversal symmetry 
must be quadratic
in the electromagnetic field.

In the fourth section, these curvature decompositions make it a simple 
matter to find a tensor
quadratic in \textbf{F} and with the symmetries of a conformal tensor. This 
defines
$\mathbf{C_F}$ up to an overall numerical factor. Specific examples can then 
be checked to find
this factor. For an electrically charged spherical black hole the value is 
$+3$. For a
magnetically charged spherical black hole the value is $- 3$. Of course 
conformally flat electro
vac solutions give a value of zero. By picking the value $+3$ the local 
gravitational field is
tuned precisely to the physical Maxwell fields. Thus the electrically 
charged spherical black
hole provides a physical example showing that $\mathbf{C_F}$ exists and is 
given by
\eqref{1.2}.

The fifth section presents this new curvature based unification. Here the 
full second Bianchi
identities require the introduction of a third rank tensor as a generalized 
Lorentz force
density that traces down to the usual Lorentz force density. The traceless 
piece of this
generalized Lorentz force density requires that a local gravitational piece 
of the current's
curvature leak out beyond the support of the currents and have the same 
support as the
electromagnetic field.

The sixth section gives the charged spherical and charged rotating black 
holes as examples of
the new unification. While a radiative null field example is not given, the 
curvature of such
expected physical solutions is characterized.

The concluding section points out some new directions that this work 
suggests.

\section{Classical Electromagnetism}

    Classical electromagnetism is a well verified theory in the limit that 
the currents and
fields are continuous matter distributions. It fails miserably with 
granularities in the field
(photons) and the currents (electrons).  Similarly, general relativity deals 
with continuous
curvature and continuous matter distributions described by the stress-energy 
tensor and cannot
deal with the granularity of matter. Nearly a century of work trying to find 
this quantum limit
of general relativity has shown that radically new ideas are required.  The 
effort
continues to be hampered by the lack of experimental results. One radical 
idea suggested
\cite{10} by general relativity is that the topology of space-time is not 
trivial, as in the
black hole solutions. In the quantum processes of emission and absorption 
the spatial support
of the energy tensor changes from connected to disconnected and vice versa. 
Although the
curvature carried by electrons and photons will be experimentally 
inaccessible for a very long
time, this curvature must be in the quantum limit of general relativity. 
While awaiting new
experimental directions, it is useful to examine classical electromagnetism 
from the
topological perspective of De Rham's co homology theory \cite{11}.

\subsection{The Maxwell Equations}

The Maxwell equations relate the electromagnetic field \textbf{F}, the 
vector potential
\textbf{A}, and the electric current \textbf{J} by
\begin{equation} \label{2.1.1}
\mathbf{F}=\mathbf{dA},  \qquad   
\mathbf{d}*\mathbf{F}=\frac{1}{\varepsilon_{0}}*\mathbf{J}.
\end{equation}
The first Maxwell equation says that the co homology class of the global 
vector potential
\textbf{A} determines the electromagnetic field \textbf{F}. This demands a 
Maxwell field to be
exact and forbids the existence of magnetic monopoles \cite{5,7,8}. The 
second says that the co
homology class of the form dual to \textbf{F} determines the electric 
current three-form
\textbf{*J}. Superficially this permits two kinds of Maxwell fields, those 
with sources where
\textbf{J} is not zero satisfying equations \eqref{2.1.1} and the source 
free Maxwell fields
obeying the source free Maxwell equations
\begin{equation} \label{2.1.2}
\mathbf{F}=\mathbf{dA}, \qquad  \mathbf{d*F}=\mathbf{0}.   \end{equation}
Topologically there are two types of source free Maxwell fields. Only 
topologically non-trivial space-times
will permit non-trivial solutions to equations \eqref{2.1.2} where 
$\mathbf{*F}$ is
closed and not exact.  There are also the trivial solutions to equations 
\eqref{2.1.2} where
$\mathbf{*F}$ is exact
\begin{equation} \label{2.1.3}
\mathbf{F}=\mathbf{dA},  \qquad \mathbf{*F}=\mathbf{dB}  \end{equation}
and \textbf{B} is a global one form.

The source free equations \eqref{2.1.2} were introduced to handle the 
radiation fields and are very
successful in the classical regime. However, all observed classical 
radiation fields obey
equations \eqref{2.1.3} and since all exact forms are closed, they trivially 
solve the source
free equations \eqref{2.1.2}. A non-trivial radiative solution to the source 
free equations would correspond
to a topological charge travelling at light speed. No charges have ever been 
observed to travel
at the speed of light and quantum theoretic arguments have been made against 
them \cite{12}.

All exact forms are closed and Poincar\'{e}'s lemma \cite{11}
gives the difference between them. This states that on any region smoothly 
contractible to a
point, all closed forms are exact. So the existence of a closed, not exact 
form is the hallmark
of a topological hole \cite{11} in space-time. Long ago it was pointed out 
\cite{10} that one
can take any space-time that has Maxwell fields with sources and commit 
space-time surgery to
remove all those regions with current. The result is a source free Maxwell 
field with
$\mathbf{*F}$ being closed and not exact in a space-time with horrible 
boundaries. The charged
black holes are an example. They are non-trivial solutions to the source 
free equations
\eqref{2.1.2} with topological charge Q found with
\begin{equation} \label{2.1.4}
Q=\varepsilon_{0}\int_S *F,    \end{equation}
where S is a two sphere enclosing the topological hole that by De Rham's 
theorem \cite{11}
belongs to the closed and not exact $\mathbf{*F}$. It was also thought 
\cite{10} that the
continuous currents might be some smooth average over topological charges, 
making the source
free equations \eqref{2.1.2} more fundamental than the Maxwell equations
\eqref{2.1.1}.

It is tempting to think that the source free equations \eqref{2.1.2} are the 
fundamental ones
with the topological charges possessing asymptotic rest frames and the 
radiative solutions
obeying equations \eqref{2.1.3} with no asymptotic rest frame. If this were 
the case and there
was some reason to believe that only a certain number of field lines could 
thread a topological
hole, then one would have the granularity of charge arising in a believable 
way. The atomicity
of charge is such a deep problem that it led to the introduction of magnetic 
monopoles
\cite{5,8} and the abandonment of the global vector potential $\A$.

\subsection{Duality Rotations}
The finishing touch on the Maxwell equations was the requirement of symmetry 
between the
source free electric and magnetic fields. Analysing the fields in a charging 
capacitor
physically demanded this aesthetic argument. On purely aesthetic grounds, 
one could imagine a
complete symmetry between electric and magnetic currents $\mathbf{J_D}$:
\begin{equation} \label{2.2.1}
\mathbf{dF_D}=\frac{1}{\varepsilon_{0}}\mathbf{*J_D}, \qquad
\qquad \mathbf{d*F_D}= \frac{1}{\varepsilon_{0}}\mathbf{*J}.  \end{equation}
These ``electromagnetic'' fields $\mathbf{F_D}$ trivially contain the 
Maxwell fields.  However,
their
non-trivial solutions are not Maxwell fields as they have no global vector 
potential $\A$.
No such magnetic currents, even the topological ones \cite{7,8}, have ever 
been observed.
The topological magnetic currents arise as non-trivial solutions to
\begin{equation} \label{2.2.2}
\mathbf{dF_D}=0, \qquad \mathbf{d*F_D}= 
\frac{1}{\varepsilon_{0}}\mathbf{*J}. \end{equation}
The magnetically charged black holes are a special case and obey
\begin{equation} \label{2.2.3}
\mathbf{dF_D}=0, \qquad \mathbf{*F_D}=-\mathbf{dA},  \end{equation}
where $\A$ is a global potential for the Maxwell field of the corresponding 
electriclaly charged
black hole.  The topological magnetic charge $\textup{Q}_{\textup{M}}$ is 
found with
\begin{equation} \label{2.2.4}
Q_M=\varepsilon_{0}\int_SF_D, \end{equation}
where S is a two sphere enclosing the topological hole that is associated 
with the closed and
not exact $\mathbf{F_D}$.

All these $\mathbf{F_D}$ come from duality rotated Maxwell fields 
\cite{1,5}. Given any two
form \textbf{F}, a duality rotated two form $e^{*\alpha}\mathbf{F}$ is 
defined by
\begin{equation} \label{2.2.5}
e^{*\alpha}\mathbf{F}=\mathbf{F}\cos \alpha +*\mathbf{F}\sin \alpha,  
\end{equation}
where $\alpha$ is a real number. Only when $\mathbf{F}$ and its dual have 
the same topology
will the duality rotated $e^{*\alpha}\mathbf{F}$ have the same topology as 
$\mathbf{F}$. The
only Maxwell fields with this property occur in equations \eqref{2.1.3} 
where $\mathbf{F}$ and
$\mathbf{*F}$ are both exact. Regardless of the differential properties of 
$\mathbf{F}$ the
duality rotation \eqref{2.2.5} maps two components from two `planes', one in 
$\mathbf{F}$ and
one in $\mathbf{*F}$, into one `plane' in $e^{*\alpha}\mathbf{F}$. On every 
quarter turn the
mapping is one to one and invertible. These quarter turns correspond to 
potential symmetries
that $\mathbf{F}$ might enjoy.

When $\alpha=\frac{\pi}{2}$, the duality rotated field 
$e^{*\alpha}\mathbf{F}$ is precisely
$\mathbf{*F}$, the dual of the field $\mathbf{F}$. The dual operation on a 
two form shuffles
space-time components into space-space components and vice versa, 
interchanging the roles of
the purely electric fields and purely magnetic fields. This interchange 
symmetry must turn
electric currents into magnetic currents and vice versa as the currents 
produce the fields.
Interchange symmetry requires the electromagnetic field to obey the field 
equations
\eqref{2.2.1}.

When $\alpha=\pi$, the new $e^{*\alpha}\mathbf{F}$ is merely the field 
$\mathbf{F}$ with a sign
reversal. This occurs when the sign of the currents or vector potentials is 
reversed.
All the electromagnetic field equations presented here enjoy this charge 
reversal symmetry.
To respect this symmetry, any quantity explicitly dependant on an 
electromagnetic field must be
quadratic in the field.

It is well known \cite{1,4,10} that any duality rotated field 
$e^{*\alpha}\mathbf{F}$ gives
rise to the same Maxwell stress-energy tensor as $\mathbf{F}$ itself in 
equation \eqref{1.1}.
Looking only at the mechanical properties $\mathbf{T_F}$, one could not tell 
them apart!
This is disastrous in general relativity where, in the electro vac case, one 
inserts
$\mathbf{T_F}$ made from a source free Maxwell field $\mathbf{F}$ into the 
Einstein equations
and solves for the metric tensor. How different is the space-time when it 
carries a
duality-rotated electromagnetic field with the same $\mathbf{T_F}$? Thinking 
physically about
the difficulty, one realizes that the gravitational fields of these 
duality-rotated
electromagnetic fields should be different.

\subsection{Requirements of Classical Mechanics}
The connection between electromagnetism and classical mechanics is made with 
the Maxwell
stress-energy tensor $\mathbf{T_F}$ given in equation \eqref{1.1}. In 
classical mechanics the
external force density on a fluid, whose mechanical properties are specified 
by a stress-energy
tensor, is given by the divergence of that stress-energy tensor. The 
electric currents also have
mechanical properties given by a stress-energy tensor $\mathbf{T_J}$ and, in 
the absence of
other external agents, its divergence is the Lorentz force density 
$\mathbf{f_L}$. This Lorentz
force density is exerted on the currents by the electromagnetic field and 
has the component
expression
\begin{equation} \label{2.3.1}
T^{\phantom{J} \mu \alpha}_{J \phantom{\J} ; \mu}=f^{\phantom{L} 
\alpha}_{L}=J^{\mu}
F^{\alpha}_{\phantom{\alpha} \mu}.
\end{equation}
Assuming the Maxwell equations, the Lorentz force law, and classical 
mechanics, an old argument
\cite{3} permits the derivation of the Maxwell stress-energy tensor. Given 
two of the three
electromagnetic items and classical mechanics, one can usually recover the 
third. Here the
argument is adapted to find the electromagnetic field equations required by 
classical
mechanics, the Maxwell stress-energy tensor and the Lorentz force law.

When no other agents are present the total stress-energy tensor \textbf{T} 
is the point wise
sum of $\mathbf{T_J}$ and $\mathbf{T_F}$
\begin{equation} \label{2.3.2}
\mathbf{T}=\mathbf{T_J}+\mathbf{T_F}. \end{equation}
With nothing else to push on the system, \textbf{T} must be divergence free 
and obey the
component equation
\begin{equation} \label{2.3.3}
T^{\mu}_{\phantom{\mu} \alpha ; \mu}=0. \end{equation}
Combining the last three equations yields
\begin{equation} \label{2.3.4}
f_{L \phantom{L} \alpha}=J^{\mu}F_{\alpha \mu}=T^{\phantom{J} \mu}_{J 
\phantom{J} \alpha ; \mu}
=-T^{\phantom{F} \mu}_{F \phantom{J} \alpha ; \mu}.
\end{equation}
Now, only using tensor identities including
\begin{equation} \label{2.3.5}
*F_{\alpha \beta}=\frac{1}{2}\eta_{\alpha \beta \mu \nu}F^{\mu \nu}, \qquad
F_{\alpha \beta}=-\frac{1}{2}\eta_{\alpha \beta \mu \nu}*F^{\mu \nu}, 
\end{equation}
\begin{equation} \begin{aligned} \label{2.3.6}
F^{\mu \nu}_{\phantom{\mu \nu} ; \nu}\eta_{\mu \alpha \beta 
\gamma}&=*F_{\alpha \beta ; \gamma}
+ *F_{\gamma \alpha ; \beta}+*F_{\beta \gamma ; \alpha}, \\
-*F^{\mu \nu}_{\phantom{\mu \nu} ; \nu}\eta_{\mu \alpha \beta \gamma}&=
F_{\alpha \beta ; \gamma} + F_{\gamma \alpha ; \beta} + F_{\beta \gamma ; 
\alpha},
\end{aligned} \end{equation}
and without using any electromagnetic field equations the divergence of 
equation \eqref{1.1}
can be beaten into
\begin{equation} \label{2.3.7}
T^{\phantom{F} \mu}_{F \phantom{F} \alpha ; \mu}=\varepsilon_{0}(F^{\mu 
\nu}_{\phantom{\mu \nu}
; \nu}F_{\mu \alpha}
+ *F^{\mu \nu}_{\phantom{\mu \nu} ; \nu}*F_{\mu \alpha}).  \end{equation}
Comparing the equations \eqref{2.3.4} and \eqref{2.3.7}, one finds
\begin{equation} \label{2.3.8}
J^{\mu}F_{\mu \alpha}=\varepsilon_{0}(F^{\mu \nu}_{\phantom{\mu \nu} ; 
\nu}F_{\mu \alpha}
+ *F^{\mu \nu}_{\phantom{\mu \nu} ; \nu}*F_{\mu \alpha}). \end{equation}
Generally, this can be true only when
\begin{equation} \label{2.3.9}
F^{\alpha \mu}_{\phantom{\mu \nu} ; 
\mu}=\frac{1}{\varepsilon_{0}}J^{\alpha},
\qquad *F^{\alpha \mu}_{\phantom{\mu \nu} ; \mu}=0. \end{equation}
Rewriting these equations in differential forms returns equations 
\eqref{2.2.2} that are not
the Maxwell equations. In the early part of the last century it was assumed 
that space-time
topology was trivial and under that assumption equations \eqref{2.3.9} do 
become the Maxwell
equations.

It could be argued that the existence of magnetic currents requires the 
Lorentz force law
to be taken as
\begin{equation} \label{2.3.10}
T^{\phantom{J} \mu}_{J \phantom{J} \alpha ; \mu}=f_{L \phantom{L} 
\alpha}=J^{\mu}F_{D
\phantom{\alpha} \alpha \mu}-J^{\phantom{D} \mu}_{D}*F_{D \phantom{\alpha} 
\alpha \mu}.
\end{equation}
Where upon the previous argument requires the field equations to be
\begin{equation} \label{2.3.11}
F^{\phantom{D} \alpha \mu}_{D \phantom{\alpha \mu} ; 
\mu}=\frac{1}{\varepsilon_{0}}J^{\alpha},
\qquad *F^{\phantom{D} \alpha \mu}_{D \phantom{\alpha \mu} ; \mu}=
-\frac{1}{\varepsilon_{0}}J^{\phantom{D} \alpha}_{D}.  \end{equation}
These give rise to the field equations \eqref{2.2.1} when written as 
differential forms.
Thus  with the modified Lorentz force density (2.3.10), classical mechanics 
can accommodate both
types of magnetic currents occurring in the field equations \eqref{2.2.2} 
and \eqref{2.2.1}.

\subsection{Null Electromagnetic Fields}
The null electromagnetic fields \cite{6} occur when the Lorentz invariants 
\eqref{1.3} vanish
everywhere. This algebraic constraint requires the electric and magnetic 
fields at any point to
be perpendicular and the magnitude of the electric field to be the speed of 
light times the
magnitude of the magnetic field. The plane spanned by the electric and 
magnetic field vectors
is perpendicular to the momentum density or Poynting vector of the 
electromagnetic field at
that point. A duality rotation of $\alpha$ merely rotates both electric and
magnetic field vectors by an angle of $\alpha$ about an axis along the 
Poynting vector, while
leaving them perpendicular and with the same magnitudes. If the null field 
were a plane
polarized electromagnetic wave, a duality rotation of $\alpha$ would rotate 
the plane of
polarization by $\alpha$. If the null field were a circularly polarized 
plane wave then the
duality rotation of $\alpha$ would give the wave a phase shift $\alpha$. 
Thus the duality angle,
which is overlooked by the Maxwell stress-energy tensor, fixes the 
polarization or phase
information of the electromagnetic field.

Null fields arise when $\F$ and $\mathbf{*F}$ share a common direction 
\textbf{L} that must be
null.  In geometric units the fields are
\begin{equation} \label{2.4.1}
\mathbf{F = L \wedge P}, \qquad  \mathbf{*F}=\Lb \wedge \mathbf{Q},   
\end{equation}
where $\Pb$ and $\Q$ must be perpendicular spacelike one forms.  This 
expression and the
four dimensionality of space-time require
\begin{equation} \begin{aligned} \label{2.4.2}
\g(\Lb,\Lb)&=\g(\Lb,\Pb)=\g(\Lb,\Q)=\g(\Pb,\Q)=0, \\
\g(\Pb,\Pb)&=\g(\Q,\Q)>0.
\end{aligned} \end{equation}
The last equality follows from the vanishing of the Lorentz invariant 
\eqref{1.3a} in the
well-known \cite{4,10} identity
\begin{equation} \label{2.4.3}
F^{\mu \alpha}F_{\mu \beta}-*F^{\mu \alpha}*F_{\mu \beta}=\frac{1}{2}
\delta^{\alpha}_{\beta}F^{\mu \nu}F_{\mu \nu}.
\end{equation}
Notice that since \textbf{L} is null, the first of equations \eqref{2.4.2} 
fixes \textbf{P}
and \textbf{Q} up to a constant multiple of \textbf{L}.

The currents are divergences of the fields and here one has
\begin{subequations}
\begin{equation} \label{2.4.4a}
\mathbf{Div\left( F\right)=\lbrack P,L\rbrack + \left( Div P\right) L - 
\left(Div L\right) P},
\end{equation}
\begin{equation} \label{2.4.b}
\mathbf{Div\left( *F\right)=\lbrack Q,L\rbrack + \left( Div Q\right) L - 
\left(Div L\right) Q}.
\end{equation} \end{subequations}
The \textbf{L, P}, and \textbf{Q} can always be chosen \cite{6} so that
\begin{equation} \label{2.4.5}
\mathbf{\lbrack P,L\rbrack = \lbrack Q,L \rbrack = 0},
\qquad \mathbf{Div L = Div P = Div Q = 0}, \end{equation}
forcing the currents to vanish.  Such null fields obey the field equations
\begin{equation} \label{2.4.6}
\mathbf{d F_D=0}, \qquad \mathbf{d*F_D=0}.
\end{equation}
So these null fields permit topological electric and magnetic currents where 
either
$\mathbf{F_D}$ or $\mathbf{*F_D}$ are closed and not exact. Like the Maxwell 
fields of
equations \eqref{2.1.3}, these $\mathbf{F_D}$ have the same topology under 
duality rotation.
The difference between the field equations \eqref{2.1.3} and \eqref{2.4.6} 
are the topological
currents permitted by equations \eqref{2.4.6}. These topological currents 
correspond to
magnetic or electric charges travelling at light speed. These unobserved 
charges can be removed
and the field equations \eqref{2.4.6} reduced to the Maxwell fields 
\eqref{2.1.3} by requiring
the null fields to have the potentials
\begin{equation}  \label{2.4.7}
\A =\lambda \Pb, \qquad  \mathbf{B} = \lambda \Q ,\end{equation}
where \textbf{P} and \textbf{Q} are closed or exact
\begin{subequations} \label{2.4.8}
\begin{equation} \label{2.4.8a}
\mathbf{dP=0}, \qquad \mathbf{dQ=0},    \end{equation}
and $\lambda$ is a function, such that
\begin{equation} \label{2.4.8b}
\mathbf{L=d} \lambda.   \end{equation}  \end{subequations}

With the currents vanishing, the divergence of the Maxwell stress-energy 
tensor also vanishes
by equation \eqref{2.3.7}. Applying this condition to equation \eqref{1.1} 
for the Maxwell
stress-energy tensor built with the null fields \eqref{2.4.1} yields the 
following additional
constraints on $\Lb$, $\Pb$, and $\Q$
\begin{equation} \label{2.4.9}
\mathbf{\nabla_L L= \nabla_L P= \nabla_L Q = 0}. \end{equation}
So $\Lb$ is tangent to a null geodesic along which $\Pb$ and $\Q$ are 
parallel.

There are some very unphysical null fields \cite{13} and one wonders how 
well they model the
physical radiation fields. By radiation fields one has in mind beams or 
pulses of light
travelling in otherwise empty space. Such objects have well measured 
mechanical properties
\cite{14,15} and should obey the source free Maxwell equations 
\eqref{2.1.3}.

\section{Space-time Curvature}
    The full Riemann curvature \textbf{R} of space-time decomposes by trace 
into a piece
\textbf{M} consisting
only of curvature traces, and the totally traceless remains \textbf{C} of 
the full curvature,
known as
Weyl's conformal tensor
\begin{equation}  \label{3.1}  \mathbf{R} = \mathbf{M} + \mathbf{C}.  
\end{equation}
The Einstein equations give the components of \textbf{M} explicitly in terms 
of the total
stress-energy
tensor of the matter present in the space-time. Only by solving the Einstein 
equations
for the metric tensor can one directly find the full
curvature \textbf{R} and, consequently, \textbf{C}. Since \textbf{C} is the 
only piece of the
curvature that survives
outside the matter distribution \textbf{M}, it merits being named  ``the 
gravitational
field'' of the
matter described by \textbf{M}. Further, when freely falling towards 
\textbf{M}, \textbf{C}
is the only one of the obvious
candidates for ``the gravitational field'' that does not adopt its flat 
space-time values.
So far
no explicit conditions have been placed on \textbf{C}. It is whatever it has 
to be to permit
a solution
of the Einstein equations. At worst it vanishes in conformally flat 
solutions; so that whatever
bizarre matter or boundary conditions obtain, this matter has no 
gravitational field. The
hallmark for the physical existence of any material thing is its 
gravitational field. This
forces consideration of the gravitational field of the electromagnetic field 
and removes
conformally flat solutions as unphysical.

\subsection{Splitting the Conformal Tensor}

The full second Bianchi identity is conveniently written as the divergence 
on the first
index of $*\mathbf{R}*$ and denoted by
       \begin{equation} \label{3.1.1}  \Div \mathbf{*R*} =  0,  
\end{equation}
where $*\mathbf{R}*$ is the double dual of the full curvature defined by
\begin{equation} \label{3.1.2}
*R*^{\alpha \beta}{}_{\gamma \delta}=\frac{1}{4}\eta^{\alpha \beta \mu \nu}
R_{\mu \nu}{}^{\rho \lambda}
\eta_{\rho \lambda \gamma \delta}.   \end{equation}
The double dual of any fourth rank tensor may be analogously defined. The 
decomposition
\eqref{3.1} and the identity \eqref{3.1.1} allows the invariant 
decomposition of the
gravitational field into
   \begin{equation}  \label{3.1.3} \mathbf{C = C_0 + C_1},    \end{equation}
where $\textbf{C}_\textbf{\scriptsize 0}$ and 
$\textbf{C}_\textbf{\scriptsize 1}$ respectively
obey
\begin{equation} \label{3.1.4}  \Div\mathbf{*C_0*} = 0,   \qquad  
\Div\mathbf{*C_1*}
=-\Div\mathbf{*M*}\not=0.
\end{equation}
The homogeneous piece $\textbf{C}_\textbf{\scriptsize 0}$ is the non-local 
part of the
gravitational field and depends on the
arrangement of distant matter containing fields, currents, and neutral 
matter. The inhomogeneous
piece $\textbf{C}_\textbf{\scriptsize 1}$ is the local gravitational field 
of the matter
\textbf{M} and has the same support
as \textbf{M}. Although algebraically and geometrically very different than 
\textbf{M},
$\textbf{C}_\textbf{\scriptsize 1}$ will contain the same arbitrary 
functions that comprise
\textbf{M}, ensuring their common support.

\subsection{Splitting the Riemann Tensor}

Decomposing the curvature \textbf{R} by trace yields three terms
\begin{equation} \label{3.2.1} \mathbf{R = M_2 + M_1 + C},    \end{equation}
where $\textbf{M}_\textbf{\scriptsize 2}$ depends on the curvature scalar R
\begin{equation} \label{3.2.2}  M^{\phantom{2} \alpha \beta}_{2 
\phantom{\alpha} \gamma \delta}=\frac{1}{12}
\delta^{\alpha \beta}_{\gamma \delta}R \end{equation}
and $\textbf{M}_\textbf{\scriptsize 1}$ depends on the trace free part of 
the Ricci tensor
\begin{equation} \label{3.2.3}
M^{\phantom{1} \alpha \beta}_{1 \phantom{\alpha}\gamma 
\delta}=-\frac{1}{2}\delta^{\alpha \beta \rho}_{\gamma \delta \lambda}
\left(R^{\lambda}_{\rho}-\frac{1}{4}\delta^{\lambda}_{\rho}R \right).
\end{equation}
This decomposition is unique in that the subscripts on 
$\textbf{M}_\textbf{\scriptsize 2}$
and $\textbf{M}_\textbf{\scriptsize 1}$ refer to the number of
non-zero traces in each term. The term \textbf{M} in equation \eqref{3.1} is 
the sum of
$\textbf{M}_\textbf{\scriptsize 2}$ and $\textbf{M}_\textbf{\scriptsize 1}$. 
The
trace of the Einstein equations determines the curvature scalar in
$\textbf{M}_\textbf{\scriptsize 2}$, and $\textbf{M}_\textbf{\scriptsize 1}$ 
is found with
the trace free part of the Einstein equations.

Just as one splits second rank tensors into symmetric and antisymmetric 
pieces. One can
define \cite{9} for fourth rank tensors
\begin{subequations} \label{3.2.4} \begin{equation} \label{3.2.4a}
\mathbf{R_{+}} =\frac{1}{2}\mathbf{( R + *R*)}, \end{equation}
   \begin{equation} \label{3.2.4b}
\mathbf{R_{-}}=\frac{1}{2}\mathbf{(R-*R*)}.\end{equation} \end{subequations}
Since this application of the Hodge star obeys $* * =-1$, one discovers 
\cite{9}
\begin{subequations} \label{3.2.5} \begin{equation} \label{3.2.5a}
\mathbf{*R_{+}*} = \mathbf{+R_{+}},
\qquad \mathbf{*R_{+}+R_{+}*} = 0, \end{equation}
\begin{equation} \label{3.2.5b} \mathbf{*R_{-}*}=\mathbf{-R_{-}},
\qquad \mathbf{*R_{-}-R_{-}*}=0.
\end{equation} \end{subequations}
Analogous relations to \eqref{3.2.4} and \eqref{3.2.5} hold for any fourth 
rank tensor. It is
well known
that \textbf{C} behaves like $\mathbf{R_{-}}$, as does 
$\textbf{M}_\textbf{\scriptsize 2}$ by
equation \eqref{3.2.2}. The traceless Ricci piece 
$\textbf{M}_\textbf{\scriptsize 1}$ behaves
like $\textbf{R}_{+}$ according to equation \eqref{3.2.3}. Using this to 
find the double dual
of equation \eqref{3.2.1} yields
\begin{equation} \label{3.2.6} \mathbf{*R*}=\mathbf{-M_2+M_1-C},  
\end{equation}
which allows the second Bianchi identity \eqref{3.1.1} to be written
\begin{equation} \label{3.2.7}  \Div\mathbf{*R*}=-\Div\mathbf{M_2+ \Div 
M_1-\Div C}=0.
\end{equation}
All the terms in the middle expression are of the form
\begin{equation} \label{3.2.8}
D^{\alpha}_{\beta \gamma}=T^{\mu \alpha}{}_{\beta \gamma ; \mu} \> ,
\end{equation}
where \textbf{D} is antisymmetric on its lower two indices. Such a third 
rank tensor can be
decomposed by trace, into a trace free piece and a piece with trace. Since 
\textbf{C} is
totally traceless, its divergence must also be traceless. Examining equation 
\eqref{3.2.2}
shows that the divergence of $\textbf{M}_\textbf{\scriptsize 2}$
can have no traceless piece. Thus it is left to the divergence of
$\textbf{M}_\textbf{\scriptsize 1}$ to provide both trace and
traceless pieces to annihilate the other divergences.

Applying this to the electromagnetic field with a traceless 
$\textbf{T}_\textbf{\scriptsize F}$,
there is only an $\textbf{M}_\textbf{\scriptsize 1}$ component for the 
mechanical
contribution of \textbf{F} to the curvature. This is the 
$\textbf{M}_\textbf{\scriptsize F}$
reported in equation \eqref{1.4}. For the second Bianchi identity 
\eqref{3.1.4} to work in general
there must also be a gravitational piece $\textbf{C}_\textbf{\scriptsize F}$ 
of the curvature
also carried by the electromagnetic field. Thus the second Bianchi identity 
and the Einstein
equations with $\textbf{T}_\textbf{\scriptsize F}$, generally require the 
existence of
the local gravitational field carried by the electromagnetic field.

\subsection{Duality Rotation of Curvature}
Duality rotation is mere index shuffling and must not be restricted to 
electromagnetic
fields. With the electromagnetic field carrying a curvature, one must 
shuffle the curvature
indices in the same way to get the curvature carried by the duality-rotated 
electromagnetic
field. Applying it to curvature would give
\begin{equation} \label{3.3.1} e^{* \alpha}\mathbf{R}=\mathbf{R}\cos\alpha+
\mathbf{*R}\sin\alpha.
\end{equation}
Since all the `planes' in \textbf{F} are duality rotated, the same should be 
done for curvature,
naturally giving
\begin{equation} \label{3.3.2}
e^{* \alpha}\mathbf{R}e^{* \alpha}= 
\mathbf{R}\cos^2\alpha+\mathbf{*R*}\sin^2{\alpha} +
\mathbf{(*R+R*)}\cos\alpha\sin\alpha.   \end{equation}
This expression simplifies considerably when the components $\mathbf{R_{+}}$ 
and
$\mbox{\textbf{R}}_{-}$ from equation \eqref{3.2.4} are used. With equation 
\eqref{3.2.5} one gets
\begin{subequations} \label{3.3.3} \begin{equation} \label{3.3.3a}
e^{* \alpha}\mathbf{R_{+}}e^{* \alpha}=\mathbf{R_{+}} \end{equation}
\begin{equation} \label{3.3.3b}
e^{* \alpha}\mathbf{R_{-}}e^{* \alpha}=e^{*2 \alpha} \mathbf{R_{-}} 
\end{equation}
\end{subequations}
So $\textbf{M}_\textbf{\scriptsize 1}$ and $\mathbf{M_F}$, behaving like
$\mbox{\textbf{R}}_{+}$, are invariant under duality rotation, but
$\textbf{M}_\textbf{\scriptsize 2}$, \textbf{C}, and 
$\textbf{C}_\textbf{\scriptsize F}$,
behaving like $\mathbf{R_{-}}$ are only invariant under a duality rotation 
of $n \pi$ with integer $n$.
One can confirm these results by inserting $e^{* \alpha}$\textbf{F} for 
\textbf{F} in equation
\eqref{1.4} for $\textbf{M}_\textbf{\scriptsize F}$, and in equation 
\eqref{1.2} for
$\textbf{C}_\textbf{\scriptsize F}$.

A duality rotation of $\pi$ corresponds to charge reversal symmetry. Thus 
the mechanical
properties, the gravitational fields and curvature of generic space-times 
will be the same for
charge reversed Maxwell fields and currents. The mechanical properties 
\textbf{M} of most
matter will fix the duality angle up to a half turn with a non-vanishing
$\textbf{M}_\textbf{\scriptsize 2}$ component, which requires a 
non-vanishing trace of the
stress-energy tensor. This luxury is unavailable with the traceless
Maxwell stress-energy tensor and so it is left to the gravitational field to 
set the duality
angle up to a half turn. Note that a quarter turn producing magnetic 
monopoles reverses the
sign of $\mathbf{C_F}$.  If magnetic monopoles existed, it would be 
impossible to
consistently define a $\mathbf{C_F}$.

\section{The Local Gravitational Field Carried by the Electromagnetic Field}
The foregoing makes the determination of $\mathbf{C_F}$ very straight 
forward. Charge reversal
symmetry requires it to be quadratic in \textbf{F} and the symmetries of a 
gravitational
field require it to
obey the analogue of equation \eqref{3.2.4b}. With $\mathbf{M_F}$ given by 
equation \eqref{1.4} a
natural candidate for $\mathbf{C_F}$ is
\begin{equation} \label{4.1}  
\mathbf{F}\otimes\mathbf{F}-\mathbf{*F}\otimes\mathbf{*F}.
\end{equation}
This expression has two symmetric blocks of antisymmetric indices. But 
unlike a gravitational
field, it can have a non-zero trace and can span 4-volume. To restore these 
indicial
symmetries, one simply removes these traces with the help of the identity 
\eqref{2.4.3} for the
Lorentz invariant \eqref{1.3a} and the 4-volume containing the Lorentz 
invariant \eqref{1.3b}
\begin{equation} \label{4.2}
\frac{1}{24} \eta_{\mu \nu \rho \lambda}(F^{\mu \nu}
F^{\rho \lambda}-*F^{\mu \nu}*F^{\rho \lambda})=\frac{1}{6}*F^{\mu \nu}
F_{\mu \nu}, \end{equation}
to get
\begin{equation} \label{4.3} \begin{aligned}
C^{\phantom{F} \alpha \beta}_{F \phantom{\beta} \gamma \delta}= & 8 \pi 
\frac{G \varepsilon_{0}}{c^4}
\left(\frac{k}{2}\right)(F^{\alpha \beta}F_{\gamma \delta}-*F^{\alpha 
\beta}*F_{\gamma \delta}\
+ \\& - \frac{1}{6} \delta^{\alpha \beta}_{\gamma \delta}F^{\mu \nu}F_{\mu 
\nu}
+ \frac{1}{6} \eta^{\alpha \beta}_{\phantom{\alpha \beta} \gamma 
\delta}*F^{\mu \nu}F_{\mu
\nu}),  \end{aligned} \end{equation}
where $k$ is an undetermined constant.  Fortunately \delt  spans no 
4-volume,
\et  has no
trace, and both obey the analogue of equation \eqref{3.2.5b}. Thus equation 
\eqref{4.3} has all
the symmetries of a gravitational field.  All attempts to determine the 
constant $k$ from the
second Bianchi identities failed.  However, an example with a Maxwell field 
will at once show
the existence of $\mathbf{C_F}$ and give the constant $k$. Amazingly this 
will smuggle in a
global vector potential, as this choice for $k$ will globally set the 
duality angle in
$\mathbf{C_F}$ to that of a Maxwell field.

The simplest example, although algebraically special, is an electrically 
charged spherical black
hole whose metric in Schwarzchild coordinates is
\begin{equation} \label{4.4} \begin{aligned}
ds^2=& 
-\left(1-2\frac{M}{r}+\frac{Q^2}{r^2}\right)dt^2+\left(1-2\frac{M}{r}+ 
\frac{Q^2}{r^2}
\right)^{-1}dr^2 \\ &+ r^2d\theta^2 + r^2\sin^2\theta d\phi^2,  
\end{aligned} \end{equation}
where M is the singularity's mass and Q its charge, both in geometric units. 
The electromagnetic
field, also in geometric units is
\begin{equation} \label{4.5}
\mathbf{F} = -\frac{Q}{r^2}dt\wedge dr.  \end{equation}
This field solves the source free Maxwell equations \eqref{2.1.2}. The fact 
that $\mathbf{*F}$
is closed and not exact gives rise \cite{10} to the topological charge Q. In 
these coordinates there
is no magnetic field and so only the first Lorentz invariant \eqref{1.3a} is 
not zero. Using the
metric \eqref{4.4} to calculate the full curvature, one finds
\begin{equation} \label{4.6}
\mathbf{R}=\mathbf{C_0} + \mathbf{M_F} + \mathbf{C_F}.     \end{equation}
Here $\mathbf{C_0}$ is divergence free and proportional to $\frac{M}{r^3}$ 
and $\mathbf{M_F}$
is found by inserting the \textbf{F} from equation \eqref{4.5} into the 
general expression \eqref{1.4}
for $\mathbf{M_F}$. Both $\mathbf{M_F}$ and $\mathbf{C_F}$ in equation 
\eqref{4.6} are
proportional to $\frac{Q^2}{r^4}$. To get agreement with this $\mathbf{C_F}$ 
and the
expression \eqref{4.3}, with expression \eqref{4.5} for \textbf{F} inserted, 
requires
\begin{equation} \label{4.7}
k=+3. \end{equation}
Putting this value of $k$ in the expression \eqref{4.3}, it becomes the 
expression for
$\mathbf{C_F}$ reported in equation \eqref{1.2}.

\section{New Unification of Relativity and Classical Electromagnetism}

The first assumption of this curvature based unification is that the total 
curvature of the electrified
matter is broken into a current piece $\mathbf{R_J}$ and a field piece 
$\mathbf{R_F}$
\begin{subequations} \label{5.1} \begin{equation} \label{5.1a}
\R=\R_{\J}+ \R_{\F}.
\end{equation}
By the Einstein equations this implies the splitting of the total 
stress-energy tensor into a
current and field piece as in equation \eqref{2.3.2}. The second assumption 
is the component
expression for the curvature carried by the classical electromagnetic field
\begin{equation} \label{5.1b} \begin{aligned}
& R^{\phantom{F} \alpha \beta}_{F \phantom{\beta} \gamma \delta} =  8\pi 
\frac{G \varepsilon_{0}}{c^4}
\times \\ &\left(2F^{\alpha \beta}F_{\gamma \delta}-*F^{\alpha 
\beta}*F_{\gamma \delta} - \frac{1}{4}
\delta^{\alpha \beta}_{\gamma \delta}F^{\mu \nu}F_{\mu \nu} + \frac{1}{4}
\eta^{\alpha \beta}_{\phantom{\alpha \beta} \gamma \delta}*F^{\mu \nu}F_{\mu 
\nu}\right)
\end{aligned} \end{equation}
Applying the Einstein equations to the trace of this expression returns 
equation \eqref{1.1}
for $\mathbf{T_F}$. Finally to introduce the current and permit the 
derivation of the Maxwell
equations a generalized Lorentz force $\mathbb{F}_{\mathbf{L}}$ is 
introduced. In components it is
\begin{equation} \label{5.1c} \begin{aligned}
\mathbb{F}^{\phantom{L} \alpha}_{L \phantom{L} \beta \gamma}= & \delta^{\mu 
\alpha}_{\beta \gamma}J^{\nu}
F_{\mu \nu} -J^{\alpha}F_{\beta \gamma} -\delta^{\mu \nu}_{\beta \gamma}
J_{\mu}F^{\alpha}_{\phantom{\alpha} \nu}
+ \\ & -\varepsilon_{0}\left(*F^{\mu \alpha}*F_{\beta \gamma ; \mu}+ 
\delta^{\mu \nu}_{\beta \gamma}g^{\alpha
\lambda}*F^{\rho}_{\phantom{\rho} \mu}*F_{\rho \nu ; \lambda}\right).
\end{aligned} \end{equation} \end{subequations}
Tracing this expression on $\alpha$ and $\gamma$ gives the Lorentz force 
density \eqref{2.3.1}.
This new unification contains all the ingredients for deriving the field 
equations \eqref{2.2.2}
required by classical mechanics. The reduction of a closed \textbf{F} to an 
exact one as
required by the Maxwell equation happens because the duality angle of 
$\mathbf{R_F}$ has been
set to a Maxwell field by the choice of $k$ in equation \eqref{4.7}.

The generalized Lorentz force $\mathbb{F}_\mathbf{L}$ contains all possible 
third rank tensors of the
form `current times field'. It is important to observe that the terms with 
field derivatives in
them can be non-zero outside the current distributions. This leads one to 
expect that a `local'
piece of the current's gravitational field $\mathbf{C_{J1}}$, whose 
divergence is not zero,
leaks out with the electromagnetic field into the `vacuum' beyond the 
support of $\J$. This is
required as the assumed curvature split \eqref{5.1a} into a current piece 
$\mathbf{R_J}$
and a field piece $\mathbf{R_F}$ can cut across both $\mathbf{M}$ and 
$\mathbf{C}$ in the
curvature split \eqref{3.1}.

The generalized Lorentz force $\mathbb{F}_\mathbf{L}$ was found analogously 
to deriving the Lorentz
force from the Maxwell equations, the Maxwell stress-energy tensor and 
classical mechanics.
To start one uses the full second Bianchi identity \eqref{3.1.1} applied to 
equation \eqref{5.1a}
\begin{equation} \label{5.2}
\Div\mathbf{*R*}=\Div\mathbf{*R_J* + Div *R_F*}=0. \end{equation}
In analogy with equation \eqref{2.3.4}, the generalized Lorentz force 
$\mathbb{F}_\mathbf{L}$
is defined by
\begin{equation} \label{5.3}
8\pi\frac{G}{c^4}\mathbb{F}_{\Lb}=\mathbf{Div*R_J*}=-\mathbf{Div*R_F*}.
\end{equation}
Remembering that the components $\mathbf{M_F}$ and $\mathbf{C_F}$ behave 
respectively like the
expression \eqref{3.2.5a} and \eqref{3.2.5b}, this last equation becomes
\begin{equation} \label{5.4}
8\pi\frac{G}{c^4}\mathbb{F}_{\Lb}=-\Div\mathbf{M_F + DivC_F}. \end{equation}
The helpful identities
\begin{equation} \label{5.5} \begin{aligned}
\frac{1}{2}\eta^{\alpha \beta}_{\phantom{\alpha \beta} \gamma \delta}*F^{\mu 
\nu}F_{\mu \nu}
&=\delta^{\mu \nu}_{\gamma \delta}\left(F^{\alpha}_{\phantom{\alpha} 
\mu}F^{\beta}_{\phantom{\beta}
\nu}-*F^{\alpha}_{\phantom{\alpha} \mu}*F^{\beta}_{\phantom{\beta} 
\nu}\right) + \\ &
-\left(F^{\alpha \beta}
F_{\gamma \delta}-*F^{\alpha \beta}*F_{\gamma \delta}\right), \end{aligned} 
\end{equation}
and
\begin{equation} \label{5.6}
\frac{1}{4}\delta^{\alpha \beta}_{\gamma \delta}F^{\mu \nu}F_{\mu 
\nu}=-\frac{1}{2}
\delta^{\alpha \mu}_{\gamma \delta}\left(F^{\nu \beta}F_{\mu \nu}-*F^{\nu 
\beta}*F_{\mu \nu},
\right)
\end{equation}
are inserted into $\mathbf{C_F}$ in equation \eqref{1.2}. This 
$\mathbf{C_F}$, and $\mathbf{M_F}$
from equation \eqref{1.4} are put into equation \eqref{5.4}. This allows the 
right hand side of
equation \eqref{5.4} to be beaten into divergences of $\F$ and $\mathbf{*F}$ 
and the field
derivatives in \eqref{5.1c}. This calculation also involves the identities 
\eqref{2.3.5} and
\eqref{2.3.6}. Finally substituting the currents for the divergences 
according to equation
\eqref{2.3.9}, then gives the expression reported in assumption 
\eqref{5.1c}.

\section{Examples}
Here the charged black holes are presented as examples of the unification 
\eqref{5.1}. Both of
these black hole solutions are examples of electro vac universes with 
topological electric
currents arising from the fact that $\mathbf{*F}$ is closed and not exact.

Although no example for physical radiative null fields is given, the 
curvature of such expected
solutions is characterized with the new unification.

\subsection{$\mathbf{F^{\mu \nu}F_{\mu \nu}\not=0}$, $\mathbf{*F^{\mu 
\nu}F_{\mu \nu}\not=0}$}
A formidable physical example is the rotating charged black hole whose 
metric in Boyer
Lindquist coordinates and geometrised units is
\begin{equation} \begin{aligned} \label{6.1.1}
ds^{2}=&-\left(1-\frac{2Mr-Q^2}{r^2+u^2}\right)dt^2 -
2\frac{\left(a^2-u^2\right)\left(2Mr-Q^2\right)}{a\left(r^2+u^2\right)}d\phi 
dt
\\ &+ \frac{1}{a^2}\left(a^2-u^2\right)\left(r^2+a^2 +
\frac{\left(a^2-u^2\right)\left(2Mr-Q^2\right)}{r^2 + u^2}\right)d\phi^2 + 
\\ &
+ \frac{r^2 + u^2}{r^2-2Mr + a^2 + Q^2}dr^2  + \frac{r^2 +u^2}{a^2-u^2}du^2,
\end{aligned}
\end{equation}
where M is the mass, a is the angular momentum per unit mass, Q is the 
charge, and
$u=r\cos\theta$.  The electromagnetic field for this space-time is
\begin{equation} \label{6.1.2} \begin{aligned}
\F= & Q\frac{r^2-u^2}{\left(r^2 + u^2\right)^2}dr\wedge dt + 
2Q\frac{ru}{\left(r^2+u^2\right)^2}
du\wedge dt + \\ 
&-Q\frac{\left(a^2-u^2\right)\left(r^2-u^2\right)}{a\left(r^2 + 
u^2\right)^2}
dr\wedge d\phi -2Q\frac{ru}{a\left(r^2 + u^2\right)^2}du\wedge d\phi.  
\hspace{0.5in}
\end{aligned} \end{equation}
Although this space-time is still algebraically special, both Lorentz 
invariants \eqref{1.3}
are not zero here. With $\mathbf{R_F}$ found from inserting equation 
\eqref{6.1.2} into
equation \eqref{5.1b},
the full curvature is found to be
\begin{equation} \label{6.1.3}
\mathbf{R=C_{0}+C_{J1}+R_{F}}, \end{equation}
where this $\mathbf{C_0}$ is proportional to M, $\mathbf{C_{J1}}$ and 
$\mathbf{R_F}$
are proportional to $\textup{Q}^2$. The gravitational field $\mathbf{C_0}$ 
has terms with
factors that contain $\textup{Q}^2$ and if Q is set to zero, then 
$\mathbf{C_0}$
becomes exactly the curvature for an uncharged rotating black hole. The 
second Bianchi identity
for this space-time breaks into
\begin{equation} \label{6.1.4} \begin{aligned}
\mathbf{Div*C_{0}*}= & \mathbf{0}, \\
8\pi\frac{G}{c^4}\mathbb{F}_\mathbf{L}=\mathbf{-Div*R_F*}= & 
\mathbf{Div*C_{J1}*\not=0}.
\end{aligned} \end{equation}
The divergence free $\mathbf{C_0}$ is a non-local gravitational field 
proportional to
M. Each of the three non-zero terms can be calculated independently, showing 
their
equality. Although the Lorentz force for this space-time vanishes, the 
generalized Lorentz
force does not by virtue of the field terms in assumption \eqref{5.1c}. The 
local gravitational
field $\mathbf{C_{J1}}$ is the gravitational field of the topological charge 
and the last
equation shows that it is required to sustain the curvature of the 
electromagnetic field.
Thus the charged rotating black hole is an example of the new unification 
with a non-zero
generalized Lorentz force.

\subsection{$\mathbf{F^{\mu \nu}F_{\mu \nu}\not=0}$, $\mathbf{*F^{\mu 
\nu}F_{\mu \nu}=0}$}
The charged spherical black hole of equation \eqref{4.5} is a physical 
example of this
unification, but the spherical symmetry is so restrictive that the fields 
are essentially cut
off from the sources. The generalized Lorentz force vanishes for this 
space-time. Calculating
it explicitly with equation \eqref{5.1c} and \eqref{4.6}, one finds that the 
two field
derivatives in equation \eqref{5.1c} cancel each other. As in equation 
\eqref{3.1.4} for the
splitting of the gravitational field into non-local and local pieces, one 
finds
\begin{equation} \label{6.2.1}
\mathbf{Div*C_0*=0}, \qquad \mathbf{Div*C_F*=Div*M_F*\not= 0}.  
\end{equation}
The non-local piece $\mathbf{C_0}$ is proportional to the mass of the black 
hole and is
exactly Weyl's conformal tensor for an uncharged spherical black hole.

\subsection{$\mathbf{F^{\mu \nu}F_{\mu \nu}=0}$, $\mathbf{*F^{\mu \nu}F_{\mu 
\nu}=0}$}
Presumably null fields describe the physical radiation fields that are 
pulses or beams of
electromagnetic radiation. The principle of mass energy equivalence and 
Newtonian gravitation
requires the existence of tidal forces outside the light beam. Any such 
gravitational effects
beyond the support of the beam require a non-local gravitational field 
$\mathbf{C_0}$. The
radiation fields must possess a non–zero non-local gravitational field,
\begin{equation} \label{6.3.1}
\mathbf{C_0\not= 0}. \end{equation}
Inside the beam described by null fields the gravitational field of the 
electromagnetic
field reduces to
\begin{equation} \label{6.3.2}
\mathbf{C_F}=8\pi\frac{G\varepsilon_0}{c^4}\left(\frac{3}{2}\lbrace 
\left(\mathbf{L}
\wedge \mathbf{P}\right)
\otimes \left(\mathbf{L \wedge P} \right) - \left(\mathbf{L\wedge Q} \right)
\otimes \left(\mathbf{L \wedge Q} \right) \rbrace \right). \end{equation}
There is a large body of literature on the gravitational properties of 
light, but to the
author's knowledge there are no exact solutions to the Einstein-Maxwell 
equations found so far
that satisfy equations \eqref{6.3.1} and \eqref{6.3.2}. Without a known 
physical solution, one
can still use the unification \eqref{5.1} to characterize the expected 
physical solutions.

Even though the unification \eqref{5.1} gives back the Maxwell equations, 
there is still the problem
of massless charges \cite{12}.  Some assumption is required to reduce the 
full Maxwell
equations \eqref{2.1.1} to the Maxwell equations \eqref{2.1.3} for 
radiation. Null fields
obeying the differential relations \eqref{2.4.8} supply the two vector 
potentials required for
equations \eqref{2.1.3}. An $\F$ and $\mathbf{*F}$ satisfying these 
equations can be duality
rotated into each other and so one should expect symmetry between \textbf{P} 
and \textbf{Q}.
This is the case with equations \eqref{2.4.2}, \eqref{2.4.5} and 
\eqref{2.4.9} that the source
free null fields must obey.

However, the duality angle of $\mathbf{R_F}$ is fixed up to a half turn with 
$\mathbf{C_F}$
from equation \eqref{6.3.2} and both show no such symmetry. Inserting the 
null field \eqref{2.4.1}
into equation \eqref{5.1b} gives
\begin{equation} \label{6.3.3}
\mathbf{R_F}=8\pi\frac{G\varepsilon_0}{c^4} \lbrace 2 \left(\mathbf{L}
\wedge \mathbf{P}\right)
\otimes \left(\mathbf{L \wedge P} \right) - \left(\mathbf{L\wedge Q} \right)
\otimes \left(\mathbf{L \wedge Q} \right) \rbrace.
\end{equation}
Using this and the last of equation \eqref{5.3} to calculate the generalized 
Lorentz force
density, one finds
\begin{equation} \label{6.3.4}
\mathbb{F}^{\alpha}_{L \phantom{\alpha} \beta \gamma}=-\varepsilon_{0}
\delta^{\mu \nu}_{\beta \gamma}L^{\alpha}L_{\mu}\left(P_{\nu ; 
\lambda}P^{\lambda}
+ 2Q_{\nu ; \lambda}Q^{\lambda}\right)
\end{equation}
on applying the constraints \eqref{2.4.5} and \eqref{2.4.9}. One would 
expect the radiation
fields to be completely cut off from their sources. This requires setting 
$\mathbf{R_J}$ to
zero. By the first of equation \eqref{5.3}, the vanishing of $\mathbf{R_J}$ 
forces the
generalized Lorentz force density $\eqref{6.3.4}$ to be zero. Respecting the 
symmetry between
\textbf{P} and \textbf{Q}, that will happen when they are tangent to 
geodesics
\begin{equation} \label{6.3.5}
\mathbf{\nabla_P P=\nabla_Q Q= 0}.   \end{equation}

Remembering the non-local gravitational field \eqref{6.3.1}, the full 
curvature for a null
radiative space-time is
\begin{equation} \label{6.3.6}
\mathbf{R}=8\pi\frac{G\varepsilon_0}{c^4} \lbrace 2 \left(\mathbf{L}
\wedge \mathbf{P}\right)
\otimes \left(\mathbf{L \wedge P} \right) - \left(\mathbf{L\wedge Q} \right)
\otimes \left(\mathbf{L \wedge Q} \right) \rbrace + \mathbf{C_0},
\end{equation}
with \textbf{L}, \textbf{P} and \textbf{Q} satisfying equations 
\eqref{2.4.2}, \eqref{2.4.5},
\eqref{2.4.8}, \eqref{2.4.9}, and \eqref{6.3.5}.

Further classical mechanics and classical electromagnetism on a trivial 
space-time topology
cannot account for a light beam or pulse with a finite cross sectional area 
and an intrinsic
rotational angular momentum \cite{16}. Essentially, the argument is that 
travelling
any internal motion is frozen out when travelling at the speed of light. The 
momentum
circulation that could produce rotational
angular momentum is prohibited and any angular momentum can only be gained 
at the price of an
infinite moment arm. The only remaining possibilities to explain finite 
light beams with
measured internal angular momentum are the introduction of quantum spin or a 
non-trivial
topology that can produce rotational angular momentum as occurs with 
rotating black holes.
Perhaps quantum spin is topological in origin and results from a closed two 
form that is not
exact.

Regarding curvature as various kinds of fluid, one can interpret the second 
Bianchi identity
as regulating their flows. The special relativistic condition of no flow at 
light speed then
suggests that the curvature for radiation fields obey
\begin{equation} \label{6.3.7}  \mathbf{DivC_F = Div M_F = 0}. 
\end{equation}
These conditions do follow from equations \eqref{2.4.5}, \eqref{2.4.9} and 
\eqref{6.3.5}.
In fact these equations permit
\begin{equation} \label{6.3.8}
\mathbf{Div\left(F\otimes F\right)=Div\left(*F\otimes *F\right)=0}. 
\end{equation}

\section{Conclusions}
This work was based on splitting the full curvature in various ways and 
applying the full
second Bianchi identity to these splits.

Mechanically the Riemann tensor decomposes into
\begin{equation} \label{7.1}
\mathbf{R=M_2 + M_1 + C_1 +C_0}. \end{equation}
This permitted the application of duality rotations to the components of the 
curvature tensor.
The $\mathbf{M_1}$ component is invariant under duality rotations, but the 
$\mathbf{M_2}$
and \textbf{C} components are only invariant under a duality half turn. Such 
a half turn on a
Maxwell field corresponds to charge reversal symmetry and so space-time 
curvature is invariant
under charge reversal symmetry. Applying the second Bianchi identity to this 
decomposition
required that the trace and traceless pieces of the divergence of 
$\mathbf{M_1}$ annihilate
those of $\mathbf{M_2}$ and $\mathbf{C_1}$ respectively, while the 
divergence of $\mathbf{C_0}$
vanished. When $\mathbf{M_1}$ has non zero trace and traceless divergences, 
then $\mathbf{M_2}$,
$\mathbf{M_1}$, and $\mathbf{C_1}$ share a common support with the total 
stress-energy tensor
\textbf{T}, while the divergence free $\mathbf{C_0}$ extends beyond the 
support of \textbf{T}.
Beyond \textbf{T}, $\mathbf{C_0}$ becomes the total curvature. Splitting 
Weyl's conformal
tensor into distant $\mathbf{C_0}$ and local $\mathbf{C_1}$ gravitational 
fields hinges more on
the question of support than of divergence. There may be applications of 
this splitting
elsewhere. When the total stress-energy tensor is made of simple functions, 
as with an ideal
fluid, there may be local gravitational fields made of the same functions, 
ensuring their
common support. Further work will tell.

Turning to electromagnetism, the radiative fields are quite different than 
the non-radiative
fields. Since the absence of magnetic charges suggested the introduction of 
the global vector
potential \textbf{A}, then the absence of lightlike electric charges should 
also suggest
modifying the Maxwell equations by the introduction of a second vector 
potential \textbf{B} for
$\mathbf{*F}$ in the radiative solutions.

The curvature split between current and field cuts across both the 
mechanical and gravitational
components giving
\begin{equation} \label{7.2}
\mathbf{R=M_{2J} + M_{1J} + C_{1J} + M_F + C_F + C_0}.   \end{equation}
This is the first assumption \eqref{5.1a} in the new curvature based 
unification of classical
electromagnetism and general relativity. The expression \eqref{5.1b}
for $\mathbf{R_F}$ is an extension of the usual Einstein equations from the 
traces of curvature
to the full curvature by specifying $\mathbf{C_F}$. For the geometry to 
contain the
polarization or phase information of the electromagnetic field, that 
information must occur in
Weyl's conformal tensor. By requiring the Einstein-Maxwell equations to 
produce a curvature
containing $\mathbf{R_F}$, instead of any curvature, one is hopefully 
limiting the
Einstein-Maxwell solutions to just the physical ones. Applying the second 
Bianchi identity to
this current and field split gives rise to the assumption \eqref{5.1c} for 
the generalized
Lorentz force $\mathbb{F}_\mathbf{L}$. The non-radiative fields will, like 
the rotating charged
black hole, in general have a non-zero $\mathbb{F}_\mathbf{L}$. Radiative 
fields, being cut off
from their sources, will have a vanishing $\mathbb{F}_\mathbf{L}$.

The trace of $\mathbb{F}_\mathbf{L}$ is the Lorentz force density and, 
outside the currents,
its traceless piece ties $\mathbf{R_F}$ to the gravitational field 
$\mathbf{C_{J1}}$ of the
currents that produce \textbf{F}. $\mathbf{C_{J1}}$ has the same support as 
\textbf{F} and
like \textbf{F} has leaked out beyond the support of the currents. The 
matter here is current
and field. So even though $\mathbf{C_{J1}}$ occurs beyond the support of the 
currents, it is
still a local field occurring within the support of the matter. The charged 
rotating black hole
shows that even topological currents carry this local gravitational field in 
addition to their
non-local gravitational fields $\mathbf{C_0}$. One wonders about the 
physical meaning of the
traceless piece of $\mathbb{F}_\mathbf{L}$.

Essentially the non-local gravitational field extends beyond the support of 
the
matter that generates it and the local gravitational field has the same 
support as the matter that produces
it. Since radiative solutions have an independent existence from the sources 
that produce them
and soon leave these sources behind, the radiative solutions are in this 
sense non-local. In a
region distant from the emitting sources the curvature \eqref{7.2} reduces 
to
\begin{equation} \label{7.3}
\mathbf{R=M_F + C_F +C_0},
\end{equation}
where $\mathbf{C_0}$ is a distant gravitational field allowing for 
gravitational effects
beyond the support of \textbf{F}. A vanishing $\mathbb{F}_\mathbf{L}$ 
requires
\begin{equation} \label{7.4}
\mathbf{Div C_F = Div M_F}.
\end{equation}
This is trivially satisfied with both divergences being zero. In this case 
$\mathbf{C_F}$ has the
same support as \textbf{F}, so it is a local field. However, being 
radiation, $\mathbf{C_F}$
has the non-local characteristic of a vanishing divergence.

From an aesthetic point of view one ought to replace the assumption 
\eqref{5.1c} for the
generalized Lorentz force $\mathbb{F}_\mathbf{L}$ with the Maxwell equations 
\eqref{2.1.1} and
be rid of the magnetic monopoles from the outset. One could then still 
produce the
$\mathbb{F}_\mathbf{L}$ by the argument given at the end of section 5. 
However, from a logical
point of view it is sweeter to show that the curvature assumptions 
\eqref{5.1} require the
Maxwell equations and have the geometry eliminate the magnetic monopoles. It 
is surprising that
seemingly local demands \eqref{5.1} can give rise to the global result that 
\textbf{F} is a
Maxwell field with a global vector potential. Although equation \eqref{1.2} 
is a local
definition of $\mathbf{C_F}$, matching the duality angle of $\mathbf{C_F}$ 
to a Maxwell field
is required everywhere and is a global demand. This theoretical success 
indicates the physical
significance of $\mathbf{C_F}$.

The major discovery of this work is the expression \eqref{1.2} for 
$\mathbf{C_F}$. The
arguments that led to that expression are quite general and should defeat 
the criticism that
$\mathbf{C_F}$ was built on algebraically special black holes and will fail 
elsewhere. It would
be useful to have a physical solution to the Einstein-Maxwell equations with 
non-zero currents
that were not overwhelmed by symmetry. Then one could extend this analysis 
into the currents
and see how the full second Bianchi identity works there. Further successful 
examples will give
knowledge and comfort; but will not prove the generality for $\mathbf{C_F}$ 
that is claimed here.
However, a single credible counterexample or the observation of a magnetic 
monopole will
vitiate this work.

The small coupling constant required by the Einstein equations,
\begin{equation} \label{7.5}
8\pi\frac{G \varepsilon_0}{c^4}=1.8382\cdot10^{-54}\mbox{Volt}^{-2},
\end{equation}
permits the superposition of electromagnetic fields. It has also led many to 
believe that the
gravitational consequences of electromagnetism are insignificant. Nothing 
could be further from
the truth. It is a matter of principle to unify classical electromagnetism 
and gravitation and
the curvature-based unification presented here allows the electromagnetic 
field to appear as an
algebraically special piece of curvature. This fulfills the nineteenth 
century speculation that
gravity and electromagnetism are both aspects of Riemann curvature.

This theory is not experimentally vacuous. The smallness of the coupling 
constant merely means
that it could be a long time before curvature detectors are sufficiently 
sensitive while
withstanding an intense electromagnetic field; or sufficiently sensitive 
over very long
distances having less intense fields. One wonders about the consequences of 
$\mathbf{C_F}$ in
the environment around very strongly magnetised neutron stars \cite{17}. 
Further, what are its
consequences in the Jacobi equation for geodesic separation that might apply 
to trans galactic
travel? When two electromagnetic fields are superposed could the interaction 
terms in the
curvature have any bearing on the problem of emission or absorption?

The physical geometry of space-time is determined by specifying the metric 
tensor or the full
curvature tensor \cite{11,18}. The Einstein equations, which link classical 
mechanics to
physical geometry, may be written as
\begin{subequations} \label{7.6} \begin{equation} \label{7.6a}
M^{\phantom{1} \alpha \beta}_{1 \phantom{1} \gamma \delta}=\frac{8 \pi 
G}{c^4}\left(-\frac{1}{2}
\delta^{\alpha \beta \lambda}_{\gamma \delta \rho}\lbrace T^{\rho}_{\lambda}
-\frac{1}{4} \delta^{\rho}_{\lambda}T\rbrace \right)
\end{equation}
and
\begin{equation} \label{7.6b}
M^{\phantom{2}\alpha \beta}_{2 \phantom{2}\gamma \delta}=\frac{8 \pi 
G}{c^4}\left(-\frac{1}{12}
\delta^{\alpha \beta}_{\gamma \delta}T\right),
\end{equation} \end{subequations}
where \textbf{T} is the total stress-energy tensor and T its trace. There is 
no
mention of Weyl's conformal tensor that would complete the specification of 
the physical
geometry.

Placing constraints on Weyl's conformal tensor is the novel feature of this 
work. Such
constraints are meant to limit the solutions to those with a physical 
gravitational field.
If the constraints are too limiting and they forbid physical solutions, then 
they will have to
be altered. Similar constraints might deal with the embarrassing number of 
Ricci flat universes,
which may or may not describe gravitational radiation. It is an open 
question whether the
Einstein equations will have to be extended to the full curvature to handle 
gravitational
radiation.

\begin{acknowledgements}
This work is dedicated to the memory of Prof. M. H. L. Pryce  (January 24, 
1913--July 24, 2003).
His friendship, encouragement and support have been invaluable over the 
years and will be sorely
missed.

It is a pleasure to thank my colleagues at U.C.F.V. for their encouragement 
and support. I am
grateful to the Joint Professional Development Committee, Scholarly Activity 
Committee, and the
Research Office at U.C.F.V. who provided some release from teaching duties 
for this work. The
Visiting Scientists Program at Simon Fraser University under Prof A. Das 
supported the early
stages of this work. I began this work while on sabbatical leave to the 
University of British
Columbia and I thank Prof. W.G. Unruh for his kind hospitality. Fruitful 
discussions with him
and Prof. M.W. Choptuik made this work possible. I am grateful to Profs. P. 
Musgrave,
D. Pollney, and K. Lake of Queen's University at Kingston for their program 
GR Tensor II, that
was used extensively here. Finally thanks to Jen Godfrey for improving and 
rendering this
manuscript into \LaTeX.
\end{acknowledgements}

\end{document}